\documentstyle[aps,prb,twocolumn]{revtex}

\begin{document}

\draft

\wideabs{

\title{Comment on ``Strain effect and the phase diagram of 
La$_{1-x}$Ba$_x$MnO$_3$ thin films''}

\author{Qingshan Yuan$^{1,2,3}$}

\address{$^1$Exp VI, Center for Electronic Correlations and 
Magnetism, Universit\"at Augsburg, 86135 Augsburg, Germany\\
$^2$Texas Center for Superconductivity and Advanced Materials, 
University of Houston, Houston, TX 77204\\
$^3$Pohl Institute of Solid State Physics, Tongji University, 
Shanghai 200092, P.R.China}

\maketitle

\begin{abstract}
Recent experiments in La$_{1-x}$Ba$_x$MnO$_3$ thin films by Zhang {\it et al.} 
{[Phys. Rev. B {\bf 64}, 184404 (2001)]}
showed that the ferromagnetic Curie temperature $T_c$ is enhanced (reduced)
by tensile (compressive) strain.
The results are regarded to be anomalous because it is generally understood by 
the authors that the tensile strain should lead to a reduction of the electron 
hopping $t$ (and thus also $T_c$) due to an elongation of the (in-plane) 
Mn-O bond length. In this comment such a general understanding of 
the strain effects is disapproved. It is argued that
the tensile strain leads primarily to a larger Mn-O-Mn bond angle 
(with the Mn-O bond length nearly unchanged), thus the pure geometric 
consideration on $t$ is already in the correct direction towards 
the experimental findings by Zhang {\it et al.} 
At the same time, an indirect strain effect under 
consideration of the dynamic electron-phonon coupling is proposed, 
which may improve the quantitative explanation.

\end{abstract}


}

Very recently, the magnetic and transport properties 
for La$_{1-x}$Ba$_x$MnO$_3$ 
$(x=0.05-0.33)$ thin films grown on SrTiO$_3$ substrate were reported 
by Zhang {\it et al.} \cite{Zhang}  Depending on the doping level $x$ 
the films may suffer tensile $(x\le 0.2)$ or compressive $(x>0.2)$ strain. 
It was found that the ferromagnetic transition temperature $T_c$ is 
enhanced (reduced) by the tensile (compressive) strain. 
The findings are regarded to be anomalous because a general understanding 
by the authors is
``{\it Tensile strain elongates the in-plane Mn-O bond length $d$, 
reducing $t_0$ (due to $t_0 \propto d^{-3.5}$) and thus $T_c$; 
in contrast, compressive strain raises $T_c$}''
(see the second paragraph in Sec. III.E).
In order to explain their unusual results, Zhang {\it et al.} proposed
a mechanism involving the orbital degree of freedom, see Ref.~\onlinecite{Kanki}
for details. The idea is that an elongation 
of the in-plane Mn-O distance due to the tensile strain helps
increase the $d_{x^2-y^2}$ component of the orbital state, thus in average the
electron hopping would be enhanced because the hopping integral 
between $d_{x^2-y^2}$ orbitals is larger than that for $d_{3z^2-r^2}$ orbitals.
As consistent results with theirs, Zhang {\it et al.} mentioned
several other experimental findings: 
in Pr$_{0.5}$Ca$_{0.5}$MnO$_3$/SrTiO$_3$ thin films the charge ordering (CO) 
becomes less stable under tensile strain;\cite{Prellier} 
and in La$_{0.67}$Ca$_{0.33}$MnO$_3$/LaAlO$_3$ thin films it is favored 
by compressive strain.\cite{Biswas} Also, the same results as found
by Zhang {\it et al.} in La$_{1-x}$Ba$_x$MnO$_3$ films were obtained by
others.\cite{Pradhan}

Although the orbital-relevant mechanism proposed by Zhang 
{\it et al.} sounds good, we disagree to their general 
understanding of the strain effects, and further question 
the reality of the proposed mechanism.
In this comment, we point out that the so called general understanding,
which is extensively adopted in the literature,
\cite{Zhang,Kanki,Pradhan,Millis,Gross} is purely a {\it mis}understanding.
In fact, a simple correction to it, although not sufficient for 
a quantitative explanation, is already in the right direction toward
the experimental findings by Zhang {\it et al.} At the same time,
another mechanism in consideration of the electron-phonon ({\it e-p}) coupling
is proposed to improve the quantitative explanation.
In the following mainly the tensile strain is discussed and the case for 
the compressive strain is the reverse. 

In thin films, the tensile strain leads to an elongation of 
the in-plane lattice parameter $a$. But it does not simply signify 
an elongation of the (in-plane) Mn-O bond length $d$. 
A possible change of the Mn-O-Mn bond angle $\theta$ was
ignored by Zhang {\it et al.}\cite{Zhang} and many others.
\cite{Pradhan,Millis,Gross}
There is a simple geometric relation between $a$ and $d$: $a=2d\sin (\theta/2)$.
Moreover, the hopping $t$ between two nearest neighbor Mn sites\cite{Supp}
is $\theta$-dependent itself, which reaches the maximum when $\theta =180^o$. 
Totally, the direct dependence of $t$ on $d$ and $\theta$ 
is given by the following empirical formula:\cite{Medarde}
\begin{equation}
t \propto d^{-3.5}\sin (\theta/2) \ .
\label{t}
\end{equation}
From Eq.~(\ref{t}), if the angle $\theta$ is assumed independent of strain, 
the elongation of the lattice parameter $a$ 
(and the corresponding $d$) will lead to a reduction of
$t$ as told by Zhang {\it et al.} Once it is noted that 
the angle $\theta$ may be modified, however,
whether the hopping $t$ (and further $T_c$)\cite{Remark}
is reduced or not by the tensile strain is {\it not} obvious
even if the length $d$ is assumed elongated. 
Thus the above general understanding is {\it not} general at all.

Actually, there seems no direct experimental evidence that the 
bond length $d$ is elongated by the tensile strain in manganite thin films. 
On the contrary, recent measurements in 
La$_{1-x}$Ca$_x$MnO$_3$ $(x=0.31,\ 0.39)$ thin films
grown on SrTiO$_3$ and LaAlO$_3$ substrates, \cite{Miniotas}
with extended x-ray absorption fine structure, showed that 
the bond length $d$ is {\it fixed} at the value for 
the bulk materials, no matter whether the films are under tensile or 
compressive strain and the thickness of them.
What is really modified is just the bond angle $\theta$. 
It becomes larger (smaller) under the tensile (compressive) strain, being
consistent with the elongation (contraction) of $a$. 
A similar result was also observed in 
La$_{0.67}$Sr$_{0.33}$MnO$_3$/LaAlO$_3$ films. \cite{Ju}
In addition, it was found in bulk materials R$_{1-x}$A$_x$MnO$_3$ 
(R=La, Pr, Nd, Y; A=Ca, Sr, Ba) with $x=0.3,\ 0.33$,\cite{Munoz} 
that changes in the average radius of the ions at (R, A) site
(or equivalently the tolerance factor) only affect the mean Mn-O-Mn bond
angle, with the mean Mn-O distance nearly unchanged. 
Based on all these findings, it is plausible to believe, 
for the current La$_{1-x}$Ba$_x$MnO$_3$ thin films,
that the primary factor modified by strain 
is the angle $\theta$. Then a conclusion, which is exactly contrary 
to the so called general understanding, that is, the hopping $t$ is enhanced 
by the tensile strain due to an enlarged $\theta$, may be naturally obtained. 
In this way, the experimental findings by Zhang {\it et al.} are not
so strange as claimed. Simultaneously, the orbital-relevant
mechanism proposed by them, which is based on the hypothetical 
elongation of $d$, is not valid.

It is interesting to further make a quantitative calculation 
based on the above argument.
When the bond length $d$ is invariant from strain, Eq.~(\ref{t})
will reduce to $t \propto a$. If we assume the 
Curie temperature $T_c \propto t$, we have $T_c \propto a$.
This means that the relative change of $T_c$, i.e., $\Delta T_c/T_c$
is in the same order as that of the in-plane lattice parameter $a$, 
i.e., $\Delta a/a$. In Ref.~\onlinecite{Zhang}, for 
La$_{1-x}$Ba$_x$MnO$_3$ $(x=0.05)$ thin films which suffer tensile strain, 
the {\it out-of}-plane lattice parameter with different 
film thickness $h$ is given:
$3.872$\r{A} for $h=85$nm and $3.862$\r{A} for $h=20$nm. The relative change
is calculated by $(3.862-3.872)/3.872\simeq -0.3\%$. Then the corresponding
relative change of $a$ should have roughly the 
same magnitude but opposite sign, i.e., $\Delta a/a \simeq +0.3\%$.
On the other hand, the Curie temperature $T_c$ increases from $130$K for
$h=85$nm to $180$K for $h=20$nm, with a relative change 
$\Delta T_c/T_c =(180-130)/130 \simeq +38 \%$.
It is clear that the change of $\theta$ (or $a$) is too small to
account for the significant change of $T_c$, although this consideration 
has already given the correct tendency of the $T_c$ variation.

From the above paragraph, it has turned out that the pure geometric 
consideration on $t$ as shown by Eq.~(\ref{t}) is not enough for a 
quantitative explanation of the experimental results in 
Ref.~\onlinecite{Zhang}, thus other cooperative mechanisms are still required.
Due to the intrinsic complexity of the concrete films,
there are various factors to influence $T_c$ such as nonstoichiometry,\cite{Zhang}
disorder,\cite{Rao} phase segregation,\cite{Bibes} etc.
Here we focus on discussing the $T_c$ variation caused by strain
since the understanding of it is indispensable in thin films.
Above we have addressed the direct effect due to strain 
on the electron hopping (and $T_c$). In the following we propose an indirect
strain effect under consideration of the dynamic {\it e-p}
coupling as a potential way to improve the quantitative explanation.
Actually, since the {\it e-p} coupling is generally important in manganites,
it is natural to ask the possible relevance of it to the change of $T_c$, 
as recently discussed by Chen {\it et al.} 
in La$_{0.9}$Sr$_{0.1}$MnO$_3$ films.\cite{Chen}
Explicitly, we consider that an electron at site $i$ 
is coupled to a local stretching mode which involves the stretching motions 
at least from the in-plane oxygen ions, for example, 
the mode $B_{2g}(1)$ with notations in 
Ref.~\onlinecite{Iliev}. The coupling is of the form: 
\begin{equation}
fc_i^{\dagger}c_i Q_i\ ,
\end{equation}
or in boson representation ($\hbar =1$): 
\begin{equation}
gc_i^{\dagger}c_i (b_i+b_i^{\dagger})
\label{e-p}
\end{equation}
with $g=f/\sqrt{2m\omega}$. 
Here $c_i,\ c_i^{\dagger}$ are fermionic operators and $b_i,\ b_i^{\dagger}$
are bosonic operators, $Q_i$ represents the normal coordinate 
describing the stretching mode considered, $f,\ g$ are coupling constants, 
$m$ is the oxygen mass, and $\omega$ is the normal vibration frequency.
In consideration of the {\it e-p} coupling (\ref{e-p}),
the bare hopping $t$ will be renormalized into the polaron hopping
\cite{Yuan}
\begin{equation}
\tilde{t}\simeq t\exp [-g^2/\omega^2]=t\exp [-f^2/(2m \omega^3)]
\label{tilde}
\end{equation}
by an extra factor. The relative change of $T_c$ now results from that
of $\tilde{t}$, which may become significant as shown below. 
To analyze the possible change of 
$\omega$ due to the modification of $\theta$, we introduce the central idea 
proposed recently by Egami and co-workers in a different context.\cite{Egami}
If the Mn-O-Mn bond is straight, the vibration of the middle oxygen, when 
compressing one Mn-O bond, will inevitably stretch the other one. 
On the other hand, if the Mn-O-Mn bond is sufficiently bent, the contraction
of one Mn-O bond can be accomodated by unbending the Mn-O-Mn triangle, 
without stretching the other Mn-O bond. Therefore,
in the former case, the oxygen feels stronger restoring force when it
vibrates, i.e., the former case corresponds to a larger effective
elastic constant $K$ or vibration frequency $\omega$. In our picture, 
the tensile strain tends to straighten the in-plane Mn-O-Mn bond,
and thus will lead to a larger $\omega$ based on the above analysis. At the
same time, the coupling constant $f$ is considered to be unaffected 
in view of the fixed $d$. Then the renormalization factor will be raised. 
In particular, due to the exponential dependence
shown in Eq.~(\ref{tilde}), a tiny hardening of $\omega$ will substantially
increase the hopping $\tilde{t}$. Therefore, it is expected that 
the quantitative explanation addressed above can be largely improved.
A full quantitative explanation will be left to the future when
more experimental data are available, e.g., the 
change of $\omega$ with strain. As the first step, the experimental refinement 
to the local atomic configuration is needed to clarify the change of $\theta$
and/or $d$ with strain for the specific La$_{1-x}$Ba$_x$MnO$_3$ films 
discussed currently.

Finally, we point out that all the above ideas indicate the potential ways 
in which the electron itineracy can be modified by strain, thus they are of
general significance for the understanding of the strain effects 
in manganite thin films. 
As mentioned above, the strain effects might not always be the dominant 
for experimental explanations in concrete films, nevertheless, they must be 
seriously studied. We mention that the (in)stability 
of the CO in the manganite films\cite{Prellier,Biswas} 
refered to in the first paragraph
can be self-consistently explained by our proposal. 

In summary, the so called general understanding of strain effects 
in manganite thin films is disapproved. It is argued that
strain primarily leads to a change of
the in-plane Mn-O-Mn bond angle, thus the pure geometric 
consideration on the electron hopping
can already give the correct tendency of the $T_c$ variation
found by Zhang {\it et al.} in La$_{1-x}$Ba$_x$MnO$_3$ 
thin films. Furthermore, in order to improve the quantitative explanation, 
an indirect strain effect under consideration of the 
dynamic electron-phonon coupling is proposed.

\bigskip
The author would like to acknowledge J. Xu, C. Schneider, T. Kopp, and
R. Y. Gu for valuable discussions and J. Zhang for 
communications. This work was financially supported by the 
Deutsche Forschungsgemeinschaft through SFB 484,
the National Natural Science Foundation of China 
(Grant No. 19904007) and the Texas Center for Superconductivity and
Advanced Materials at the University of Houston.


\begin{references}
\bibitem{Supp} In the following, the in-plane hopping is simply discussed,
 which contributes the dominant to the mean hopping averaged in three directions.
\bibitem{Remark} Due to double-exchange mechanism, a larger hopping $t$ is 
 favorable to the ferromagnetism, thus leading to a larger $T_c$.
\end{references}
\end{document}